\documentclass{aastex631}
\usepackage{xcolor}

\shorttitle{Impulsive SEP Abundances}
\shortauthors{Laming \& Kuroda}

\begin{document}

\title{Element Abundances in Impulsive Solar Energetic Particle Events
}

\correspondingauthor{J. Martin Laming}
\email{j.laming@nrl.navy.mil}

\author[0000-0002-3362-7040]{J. Martin Laming}
\affil{Space Science Division, Code 7684, Naval Research Laboratory, Washington DC 20375, USA}

\author[0000-0001-9260-8555]{Natsuha Kuroda}
\affil{George Mason University, Fairfax VA 22030, USA}
\affil{Space Science Division, Code 7684, Naval Research Laboratory, Washington DC 20375, USA}



\begin{abstract}
We outline and discuss a model for the enhanced abundances of trans-Fe elements in impulsive Solar Energetic Particle (SEP) events, where large mass
dependent abundance enhancements are frequently seen. It comes
about as a variation of the ponderomotive force model for the First Ionization Potential (FIP) Effect, i.e. the increase in coronal abundance of elements like Fe, Mg, and Si that are ionized in the solar chromosphere relative to those that are neutral. In this way, the fractionation region is placed in the chromosphere, 
and is connected to the solar envelope allowing the huge abundance variations to occur, that might otherwise be problematic with a coronal fractionation
site. The principal mechanism behind the mass-independent FIP fractionation becoming the mass dependent impulsive SEP fractionation is the suppression
of acoustic waves in the chromosphere. The ponderomotive force causing the fractionation must be due to torsional Alfv\'en waves, which couple much less effectively to slow modes than do shear waves, and upward propagating acoustic waves deriving from photospheric convection must be effectively mode 
converted to fast modes at the chromospheric layer where Alfv\'en and sound speeds are equal, and subsequently totally internally reflected. We further 
discuss observations of the environments thought to be the source of impulsive SEPs, and the extent to which the real Sun might meet these conditions.

\end{abstract}



\section{Introduction} \label{sec:intro}
The detection of enhanced abundances of the rare isotope $^3$He in various regimes of solar energetic particles has come to be recognized as almost ubiquitous  throughout the solar cycle 
\citep{wiedenbeck05}. These enhancements are thought to originate in solar $^3$He rich flares leading to impulsive SEP events, followed by further acceleration
by shocks. The production of copious amounts of $^3$He by these events is one of the most remarkable abundance anomalies known in the
solar corona, with abundance enhancements relative to $^4$He on occasion of over $10^4$. Such anomalies are also often accompanied by
enhancement in abundance of trans-Fe elements \citep[e.g.][]{mason04,mason07}, which have proved equally difficult to explain. In this paper we suggest
an origin for these trans-Fe element abundances, as a special case of element abundance fractionation by the ponderomotive force that in
usual conditions gives rise to the by-now well known First Ionization Potential (FIP) effect. 

Dating back to \citet{pottasch63}, an enhancement in coronal abundance of elements that are predominantly ionized in the solar chromosphere,
by a factor of about 3-4, has been observed relative to those that are neutral. This appears for elements with First Ionization Potential (FIP) below about 10 eV, i.e. those elements with neutral atoms capable of being photoionized by H I Lyman-$\alpha$ like Fe, Mg, Si, etc.
The FIP effect was considered controversial for many years, but began to be taken seriously in the 1980's following the influential reviews of  
\citet{1985ApJS...57..151M} and \citet{1985ApJS...57..173M}. It appears to be strongest in solar active regions and coronal mass ejections. It is common
in the slow solar wind but is weaker in fast solar wind and its coronal hole source regions. It is also found in gradual SEP events.

These and other abundance anomalies (e.g. the ``Inverse'' FIP effect) have all been explained by a model invoking the ponderomotive force as the
agent of ion-neutral separation in the chromosphere 
\citep{laming04,laming09,laming12,rakowski12,laming15,dahlburg16,laming17,laming19,kuroda20,reville21}. Alfv\'en waves reflecting and 
refracting during their propagation through the chromosphere exert a force on those particles that contribute to the dielectric tensor, i.e. the electrons and ions, 
with neutrals remaining unaffected. This is an analog in magnetohydrodynamics (MHD) of work in optical physics in the trapping of atoms and molecules in
laser beams \citep[e.g.][]{ashkin70,ashkin86}, recognized with Nobel Prizes for Steven Chu (in 1997) and Arthur Ashkin (in 2018). Under what
might be considered ``usual'' chromospheric conditions \citep[e.g.][]{avrett08,heggland11,carlsson15}, an approximately mass independent fractionation of chromospheric ions results, with a 
chromospheric Alfv\'en wave amplitude in the range 5-10 km s$^{-1}$ giving the observed magnitude of fractionation.

SEPs in impulsive events show a different, mass-dependent, fractionation.  These
are events showing strong abundance enhancements in $^3$He (not FIP related), and also mass dependent abundance enhancements of heavy and
ultra-heavy ions \citep{reames00} that we will argue below are another manifestation of ponderomotive FIP fractionation. 
Previous explanations in terms of stochastic acceleration
\citep{eichler14} or reconnection \citep{drake09} place all the fractionation in the corona. This means that in extreme cases
of fractionation, {\em all} the ultra-heavy ions in a coronal loop must end up as SEPs. In this paper we investigate an extension of the ponderomotive 
force model in the chromosphere to explain these fractionations. Moving the site of fractionation to the chromosphere avoids the
``numbers'' problem, since the corona is now connected to the entire solar envelope (i.e. photosphere and convection zone) 
from which to draw its particles. The main departure from 
previous FIP models is that sound waves that otherwise would be propagating upwards from the convection zone into the chromosphere need to be 
absent. This most likely occurs through mode conversion to magnetosonic waves at the equipartition layer (plasma $\beta = 8\pi nk_{\rm B}T/B^2 = 6/5$
for $\gamma = 5/3$ gas) where sound and Alfv\'en speeds are equal. This leaves a mass dependent fractionation that can match observations of 
trans-Fe impulsive SEPs. 

Such a mechanism of fractionation has implications for the sources and environments in which impulsive SEPs are produced. These seem to be 
small active regions close to low latitude coronal holes \citep{wang06}, also producing jets as open and closed field lines exchange footpoints by magnetic reconnection.
\citet{nitta06} find EUV and X-ray brightenings coincident with impulsive SEPs events and type III radio bursts, again suggestive of an interaction
occurring on an open field line. More recently \citet{bucik21a} use Solar Orbiter to observe several $^3$He rich events. The ion injections and associated 
type III radio bursts are observed to coincide with with EUV jets and brightenings, with origins in two large complex active regions. \citet{bucik21b}
identify the source regions of impulsive SEP events with the Solar Dynamics Observatory, and measure the source temperatures and differential emission
measures. The temperatures are consistent with those inferred from the charge states of the SEPs themselves. With the exception of $^3$He/$^4$He,
other element abundance ratios are relatively insensitive to the source region temperature. Based on studies of ion cyclotron resonance heating in tokamaks, 
\citet{kazakov17,kazakov21} speculate that a connection between the $^4$He/H abundance ratio in the thermal plasma and the efficiency of 
$^3$He heating by ion cyclotron wave might exist. We take up the suggestion here that impulsive SEPs are
accelerated out of the thermal pool of particles from the observed charge states and apply it to abundances, arguing that the anomalous abundances are not
solely due to acceleration effects, and must be present in the supply of particles to the acceleration process. 

Before proceeding to this model, we first describe the ponderomotive force model in a little more detail and outline some
recent updates and improvements. Following that section 3 introduces the new ideas concerning impulsive SEP abundances, and gives sample 
calculations to compare with observations. Section 4 gives some further discussion and conclusions.

\section{Fractionation by the Pondermotive Force} \label{sec:pond}
The ponderomotive force provides an elegant means of ion-neutral separation in the chromosphere, but implementing such a process within a model
solar atmosphere inevitably leads to some inelegance. We use a heuristic model of the chromosphere \citep{avrett08} coupled with a force-free magnetic field model \citep{athay81}. The temperature, mass density and electron number density profiles agree with ``average'' values coming from time-dependent
simulations \citep{carlsson02,carlsson16}. Ionization of other elements is calculated here using the local temperature, density,
and radiation field, comprising coronal radiation from above, taken from
\citet{vernazza78}, absorbed progressively in the chromosphere by neutral H, and trapped
chromospheric Lyman $\alpha$ photons.
Atomic data  for photoionization cross sections are taken from \citet{verner96}, and for collisional rates from \citet{mazzotta98}
\citep[with updates listed in][]{laming20}. The other significant modification is the inclusion of the effects
of non-zero electron density on the dielectronic recombination
\citep{nikolic13,nikolic18}. \citet{mason04} and \citet{mason07} give impulsive SEP abundances for trans-Fe element in the mass ranges
78 - 100, 125 - 150, and 180 - 220 amu. We represent these with elements Rb, Cs, and W, with atomic masses of 85, 133, and 184 amu respectively.
Rates for Rb and Cs are extrapolated from those for Na and K, (remember we only need the balance between neutral and singly ionized species), and those for W come from \citet{asmussen98} for collisional rates and \citet{chantler95} for the photoionization.

We solve for the Alfv\'en wave energy density and hence calculate the ponderomotive force by integrating the transport equations given by
\citet{cranmer05} and \citet{laming15} for a model corona. For closed field, we take a model loop, straightened out with the model chromosphere placed at each end. For closed coronal
loops, we make the assumption that the waves are resonant with the loop, in the sense that the wave travel time from one footpoint to the other is
an integral number of wave half-periods.  The coronal Alfv\'en waves have amplitudes in the range 30 - 100 km s$^{-1}$ (depending on
loop density) matching observations and the heating requirements for the solar corona and wind \citep{dahlburg16,reville21}. We start the calculation at the chromospheric where sound waves mode convert to fast modes, and integrate back to the other chromosphere where the ponderomotive acceleration is calculated from the resulting wave field. In open field, we start the calculation at an altitude of 500,000 km, and integrate back to the chromosphere, again
calculating the ponderomotive acceleration in the same way.
For closed loops we assume resonant waves, and
\citet{cranmer07} gives a model Alfv\'en wave spectrum for a polar coronal hole, but for equatorial coronal holes there is no theoretical guidance.
We take one five minute period Alfv\'en wave with amplitude similar to that in the closed loop to reproduce the observed FIP fractionations.
The ponderomotive acceleration, $a$,  for shear or torsional Alfv\'en waves or for fast mode waves, i.e. all polarizations except for circular polarization, 
is given by \citep[e.g.][]{laming09,laming15,laming12}
\begin{equation}
a={c^2\over 2}{\partial \over\partial z}\left(\delta E^2\over B^2\right).
\end{equation}
This expression admits FIP or Inverse FIP, depending on the sign of ${\partial\left(\delta E^2/B^2\right)/\partial z}$, but usually gives FIP effect
since $B$ is generally decreasing with height while $\delta E$ increases. It only operates on charged particles, but is otherwise independent of charge giving electrons and ions the same acceleration.

The element fractionation $f_k$ of element $k$ by the ponderomotive acceleration is given in the simplest case by the equation \citep{laming17}
\begin{equation}
f_k={\rho _k\left(z_u\right)\over\rho _k\left(z_l\right)}=\exp\left\{
\int _{z_l}^{z_u}{2\xi _ka\nu _{kn}/\left[\xi _k\nu
_{kn} +\left(1-\xi _k\right)\nu _{ki}\right]\over 2k_{\rm B}T/m_k+v_{||,osc}^2+2u_k^2}dz\right\}.
\end{equation}
It is calculated from
momentum equations for ions and neutrals in a background of protons and
neutral hydrogen between upper and lower boundaries of the fractionation region, $z_u$ and $z_l$. Typically, $z_u$ is in the corona, where
everything has become ionized. Fractionations are more sensitive to $z_l$, which is taken where the ratio of gas and magnetic 
pressures $\beta = 8\pi nk_{\rm B}T/B^2 = 1.2$. A rationale for this choice is given by \citet{laming21}, who argues that photospheric 
hydrodynamic turbulence is too strong to allow fractionation, while weaker MHD turbulence in the chromosphere has little effect.
In equation 2, $\xi _k$ is the element ionization fraction and $\nu _{ki}$ and $\nu
_{kn}$ are collision frequencies of ions and neutrals with the background gas. This is
mainly hydrogen and protons, and the collision frequencies are given by formulae in \citet{laming04}. In the denominator, 
$k_{\rm B}T/m_k \left( =v_z^2\right)$ represents the square of the element
thermal velocity along the $z$-direction, $u_k$ is the upward flow speed and
$v_{||,osc}$ is a longitudinal oscillatory speed, corresponding to upward and
downward propagating sound waves. 

The approximations outlined above give a very good match to observed FIP fractionations in the solar corona and wind \citep[see][for the latest iteration]{laming19}. More recently,
direct evidence of a connection between Alfv\'enic waves and FIP fractionation has been uncovered by
\citet{baker21}. Their Figure 5 shows
a Hinode/Extreme ultraviolet Imaging Spectrograph (Hinode/EIS) FIP bias map, inferred from the ratio of the coronal abundance ratio Si/S to the photospheric ratio. This is taken from the corona above a sunspot, and is
compared with a map of Alfv\'enic perturbation amplitude in the underlying chromosphere, derived from Ca II observations taken with the Interferometric BIdirectional Spectrometer (IBIS). A strong correlation between chromospheric regions with intense Alfv\'enic waves, and coronal regions with high FIP enhancement is indicated by magnetic field lines calculated from a Potential Field Source Surface extrapolation connecting the photospheric and coronal maps. 
We consider this to be the strongest observational indication to date that Alfv\'en(ic) waves are connected with FIP fractionation. This reinforces the idea first 
put forward by \citet{laming04}, and forms the basis of our model here.

\begin{figure}[t]
\centerline{\includegraphics[width=3.5truein]{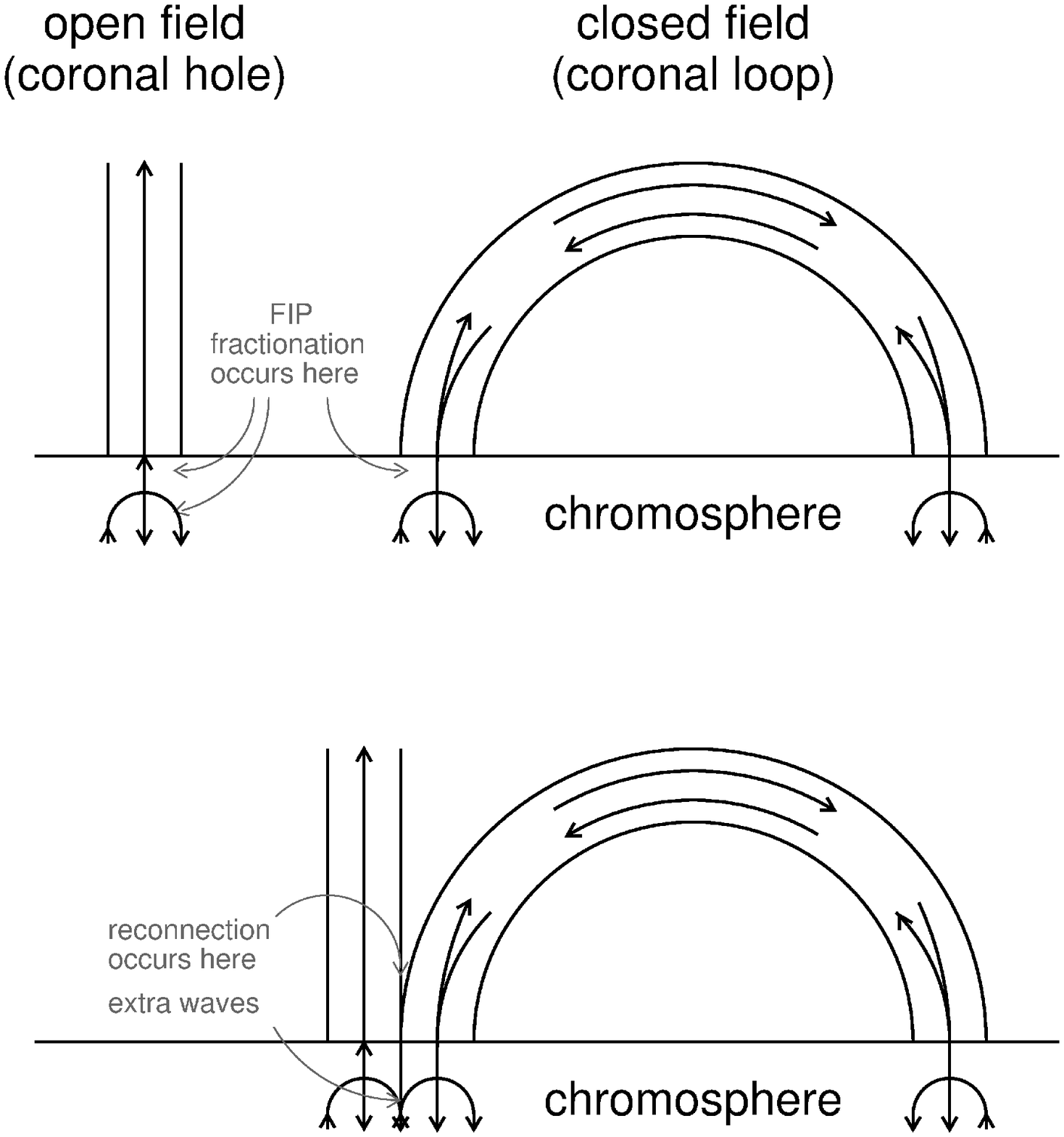}\includegraphics[width=3.5truein]{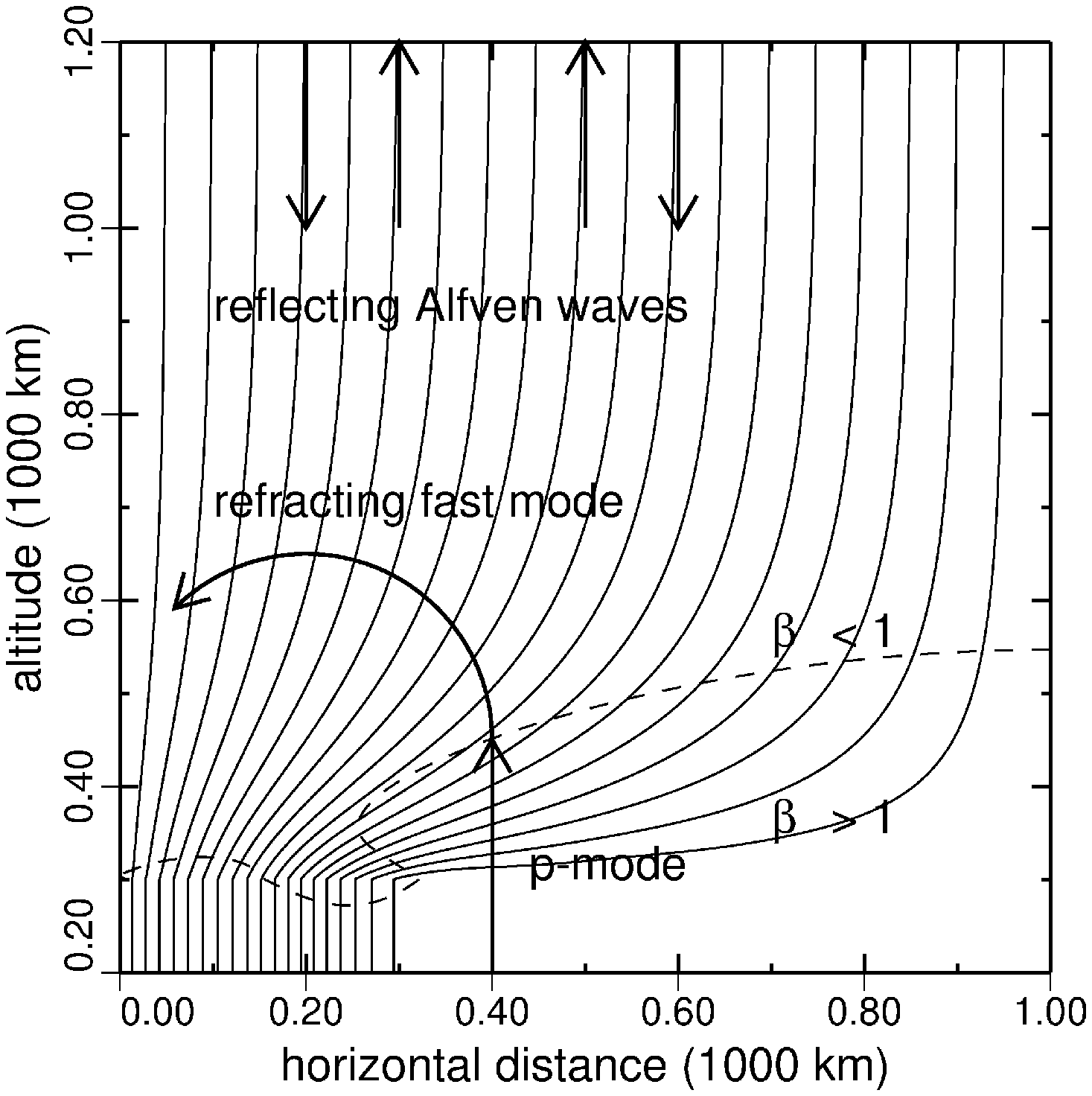}}
\caption{Left: Schematic diagrams of separate open and closed magnetic field (top) and reconnecting (bottom). Wave propagation is indicated by the black arrows. The gray arrows indicate regions of FIP fractionation and extra wave generation by the interchange reconnection in the top panel. The double arrow for the open field indicate that FIP fractionation can occur over a range of heights in the chromosphere, whereas for the closed field it is confined to the top. The gray arrows in the bottom panel indicate the regions of interchange reconnection and extra wave generation compared to the top panel.
Right: Schematic diagram of mode conversion changing upcoming acoustic waves to fast mode at $\beta = 8\pi nk_{|rm B}T/B^2 {>\atop <} 1$, and total
internal reflection of the merging fast modes. This is the key to strongly mass dependent fractionation.}
\label{fig:beta}
\end{figure}

\section{Impulsive SEP Abundances}
\subsection{Calculations}
We now turn to the problem of the mass dependent fractionation of impulsive SEPs. Following the discussion above of the likely source connected with jets,
we consider contributions from open and closed magnetic field configurations, as in the left panel of Figure \ref{fig:beta}. 
Element abundances will be a mixture of open and closed
field fractionations, occuring in the chromospheric footpoints of each field configuration as indicated by the gray arrows in the top left panel of Figure
 \ref{fig:beta}. Fractionation can occur over a greater range of chromospheric height in open field than in closed field \citep{laming19}.
Even more importantly, the reconnection (Interchange Reconnection; IR) can generate waves that propagate down to the fractionation
region in the chromosphere, as indicated by the gray arrows in the bottom panel. The fractionated plasma, when it flows upwards along the same field lines, will be injected
directly into the IR region. This is important because such abundance enhancements as we model here are not observed in the bulk solar wind.  
In this way
the fractionated plasma participates in the current sheet processes (magnetic island merging, Fermi acceleration etc) that accelerate SEPs without
encountering the plasma turbulence that accelerates the bulk solar wind. 
There is a lot of scope for detailed modeling of the IR process and the waves it produces \citep[e.g.][]{kigure10}, but
in this first attempt we restrict ourselves to modeling 5 minute period waves on the open field lines, and waves on the closed field that are resonant, in the
sense that the wave travel time from one loop footpoint to the other is an integral number of wave half periods. Since the waves do not reflect from a single
monolithic barrier as in an optical resonant cavity, but instead reflect from a range of chromospheric heights where the Alfv\'en speed is changing, we evaluate
the net ponderomotive force as an incoherent sum of the ponderomotive forces produced by the various pairs of forward and backward going waves of the same frequency in the system.

How does mass dependent fractionation occur? Equation 2 clearly admits
mass dependent fractionation if the denominator of the integrand is dominated by the term in the ion thermal speed, $2k_{\rm B}T/m_k$. The upflow velocity
term, $2u_k^2$, can be small in a scenario where the reconnection and fractionation is episodic. 
A burst of Alfv\'en waves from the reconnection travel down field lines to the
chromosphere  and fractionate the plasma, before the electron heat conduction front arrives to cause the heating and evaporation. Thus the fractionation occurs
while the plasma is at rest before being evaporated. This requires that the
Alfv\'en speed be greater than the electron thermal speed, or that the magnetic field be of order 100G or greater. The longitudinal oscillatory term,
$v_{||,osc}^2$ represents sound waves in the chromosphere, either those deriving from photospheric convection and moving upwards through
the chromosphere, or those generated locally by the Alfv\'en waves through a parametric instability. The parametric generation is unavoidable (except with 
circularly polarized waves, which we consider to be unlikely in this case), but is lower for torsional Alfv\'en waves compared to shear waves
\citep{vasheghani11,laming17}. The torsional Alfv\'en wave derives from reconnection between open field and twisted coronal loop \citep[e.g.][]{kohutova20}. 
The twist is transferred to the open field, and unwinds, exciting torsional waves. Their existence requires the cylindrical symmetry of a loop structure. Shear
Alfv\'en waves are formal solutions of the linearized MHD equations for homogeneous and infinite media. They are plane polarized, and can result from a number
of MHD processes.

\begin{figure}[ht]
\centerline{\includegraphics[width=7.0truein]{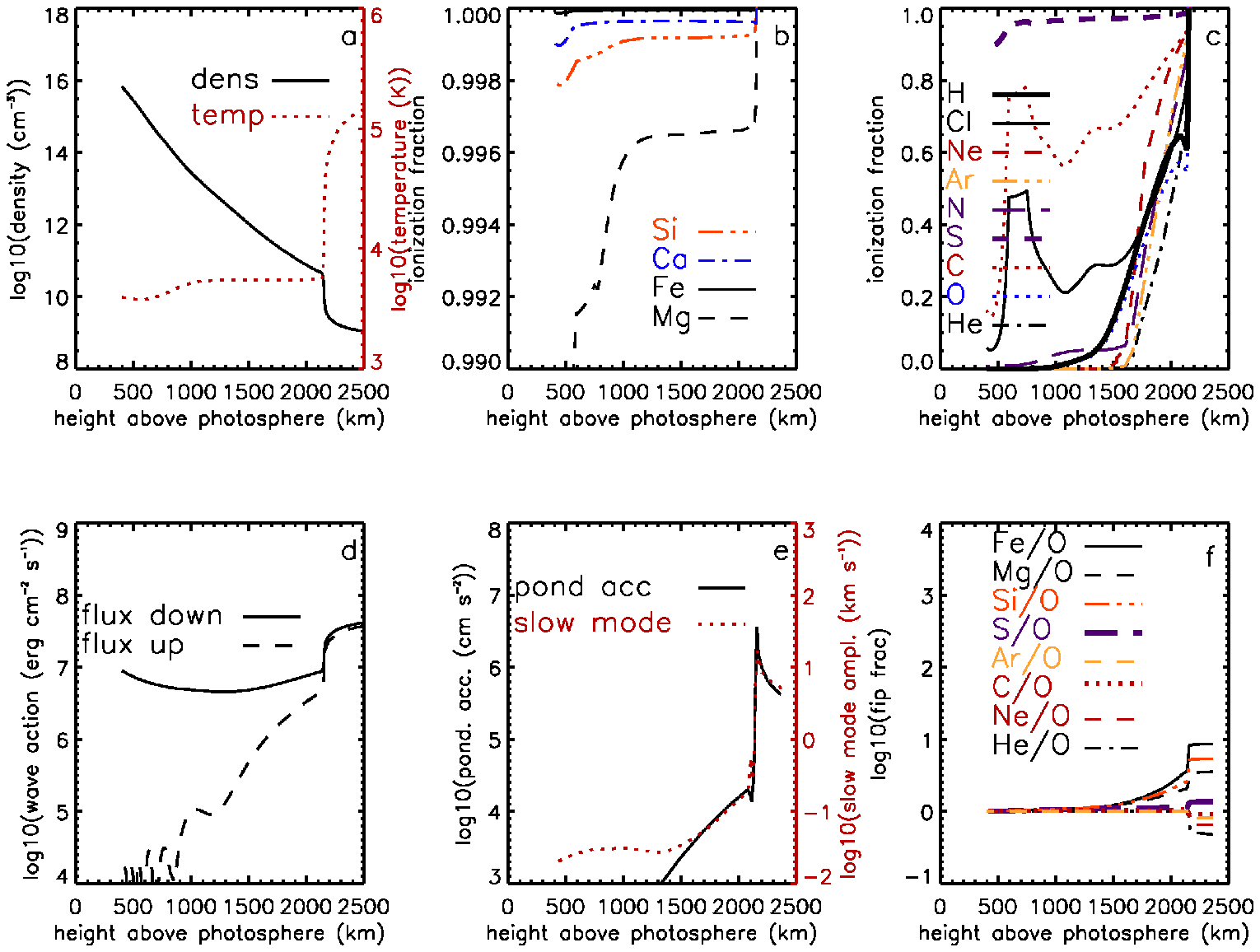}}
\centerline{\includegraphics[width=7.0truein]{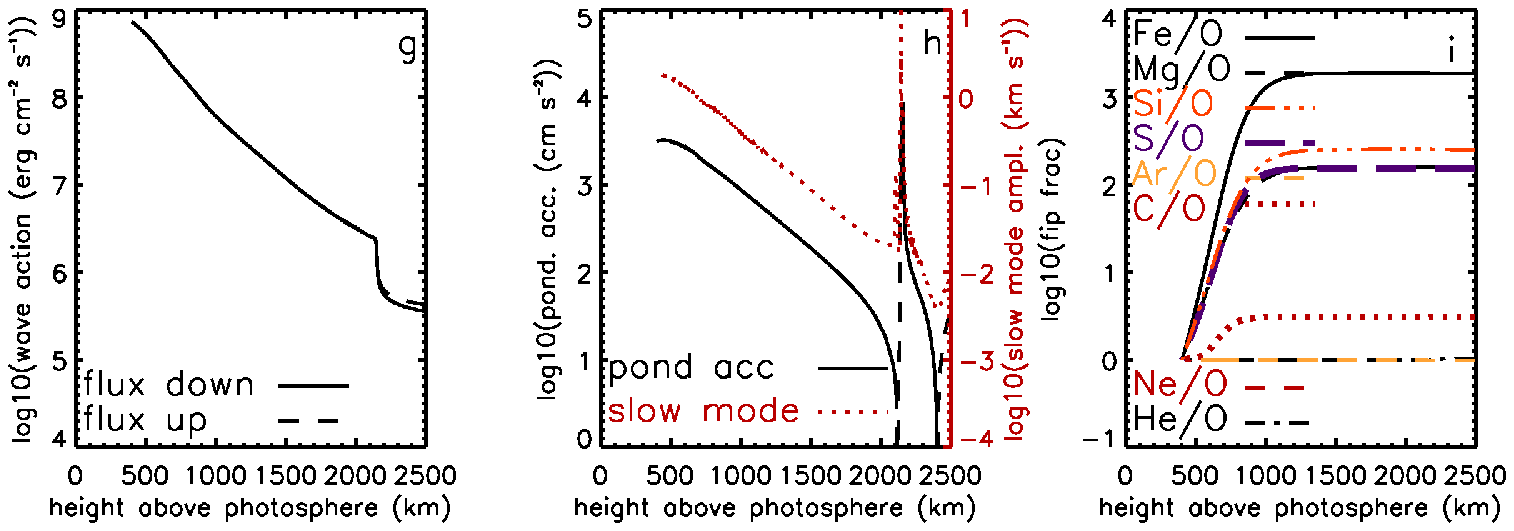}}
\caption{FIP fractionation in the chromosphere of closed and open fields. (a) density and temperature structure, (b) ionization fractions of low FIP elements
from 99\% to 100\%, (c) ionization fractions of high FIP elements (including C and S)
from 0 to 100\%, both calculated using the electron density from the \citet{avrett08} model and the same for both cases. Closed loop wave energy fluxes for upward and downward
propagating waves are given in (d), (e) shows the ponderomotive acceleration and parametrically generated slow mode wave amplitude, and (f) the FIP fractionations (on a $\log _{10}$ scale), all for model T0(h) from Table 1. Panels (g-i) repeat (d-f) but for the open field model T81(h). The difference in the location of fractionation is clear, occurring at the top of the chromosphere in a background gas of protons for the closed loop, and low down in neutral H background gas for the open field.}
\label{fig:closed}
\end{figure}

Acoustic waves moving up from the photosphere must be mode converted to fast modes at the plasma $\beta =1.2$ layer, which are then subsequently
totally internally reflected. In the absence of mode conversion,
chromospheric acoustic waves are introduced to match simulations and data
analysis in \citet{heggland11} and \citet{carlsson15}, and are added in quadrature to the slow mode waves
generated parametrically by the Alfv\'en wave driver \citep{laming12,laming15}. With mode conversion included, the same energy flux is introduced and allowed to mode convert according to the geometry, as illustrated in the
right panel of Fig. \ref{fig:beta}. The schematic shows magnetic fields \citep[calculated from][]{athay81} and wave propagation in the solar 
chromosphere, indicating the key components. Upcoming p-modes mode convert at the layer where sound and Alfv\'en speeds are equal 
($\beta = 6/5$ in $\gamma =5/3$ gas) and propagate into the $\beta < 1$ region as fast mode waves, where they undergo a total internal reflection.
The energy transmission coefficient at $\beta =6/5$ is \citep{schunker06}
\begin{equation}
T=\exp -{\pi H_Dk\sec\theta\sin ^2\theta}\sim\exp -1.6\sec\theta\sin ^2\theta
\end{equation}
where $H_D\simeq 150$ km is the density scale height, $k$ is the wavevector and $\theta$ is the angle between ${\bf k}$ and ${\bf B}$, also known as the 
``attack'' angle. The numerical
expression is evaluated for 5 minute waves and an Alfv\'en speed at $\beta =1$ of 6 km s$^{-1}$, leading to about 30\% sound wave transmission when 
$\theta = 45^{\circ}$. Higher frequency waves are more strongly mode converted. The remaining 70\% of the energy is mode converted to fast mode waves which are totally internally reflected close to $\beta=1$.

Figure \ref{fig:closed} illustrates the chromospheric portion of calculations for closed and open fields. The closed field, 
Panel (a) shows the density and temperature structure, 
(b) the ionization fractions of low-FIP elements and (c) those for high-FIP elements (including C and S) , which are all the same for open and closed field
regions. For the closed loop, the Alfv\'en waves are assumed to be resonant with the
loop. The chromospheric wave field shown in (d) has the characteristic pattern of unbalanced waves, with more flux going down than coming up. 
The ponderomotive acceleration shown in (e) is strongly spiked at the top of the chromosphere where the density gradient in (a) is strongest. The parametric
slow modes excited by the Alfv\'en wave driver are also strongest in this region. Finally (f) shows the resulting fractionations, with maximum abundance
enhancement happening at the top of the chromosphere where the ponderomotive acceleration is strongest.

Panels (g-i) the same set of panels (d-f), but here for the chromospheric portion of a calculation for an open field structure. The wave attack
angle at the $\beta = 1.2$ layer is 81.4$^{\circ}$ (model T81(h) in Table 1, designated `(h)' to reflect the higher initial wave amplitude for the hybrid model application), meaning that most of the upcoming slow modes are mode converted to fast modes and undergo a total internal
reflection. This attack angle has been chosen to give the best match with observations shown in Fig. 3c below. In reality of course, a range of attack angles, possibly varying with time, leading to time-varying fractionation during the Impulsive SEP event could well happen. Our goal here is more modestly to establish the plausibility of such a fractionation mechanism. The upgoing and downgoing wave energy fluxes are more balanced in panel (d).
The ponderomotive acceleration is quite different to the previous example, in that it shows positive (solid line) and negative (dashed line) regions, but most of the
fractionation happens low down just above the $\beta =1.2$ layer (which is at 400 km altitude for the model presented here). Fractionations are greatly
enhanced by the absence of photospheric p-modes in the chromosphere, and by the background gas in the low chromosphere being dominated by neutral H.

\subsection{Results}
\begin{figure}[t!]
\centerline{\includegraphics[width=3.5truein]{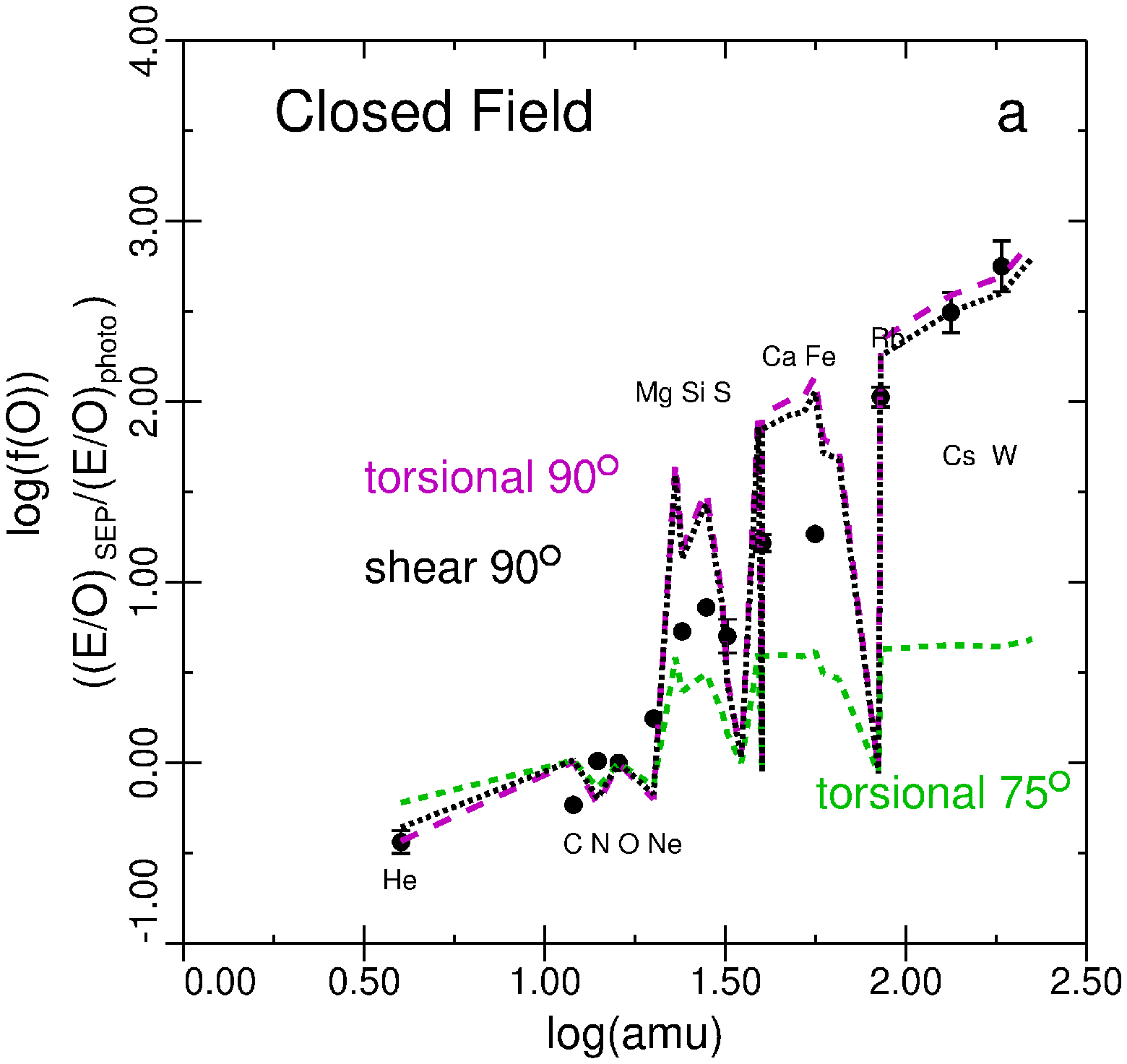}\includegraphics[width=3.5truein]{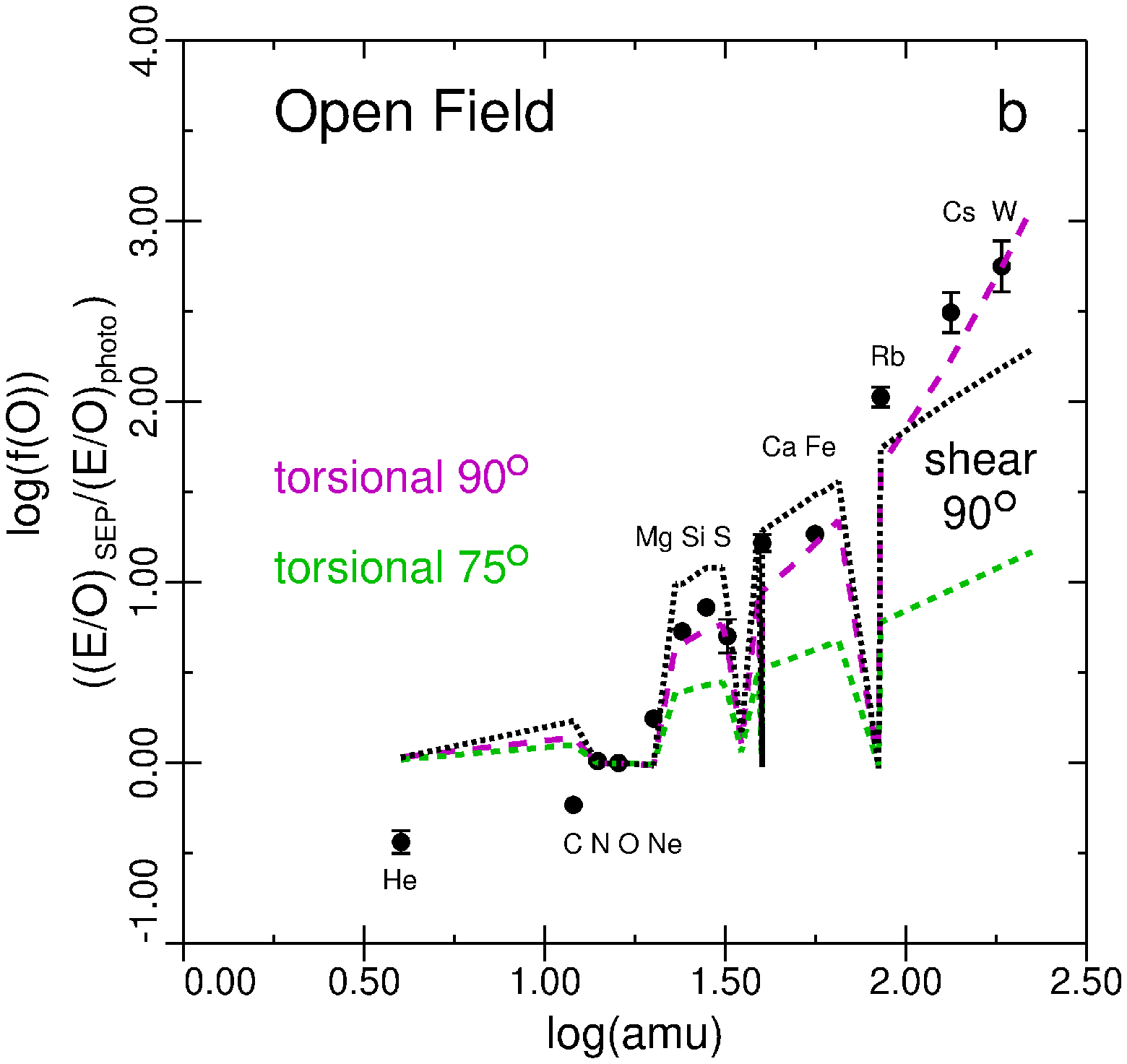}}
\vskip -0.75truein
\centerline{\includegraphics[width=3.5truein]{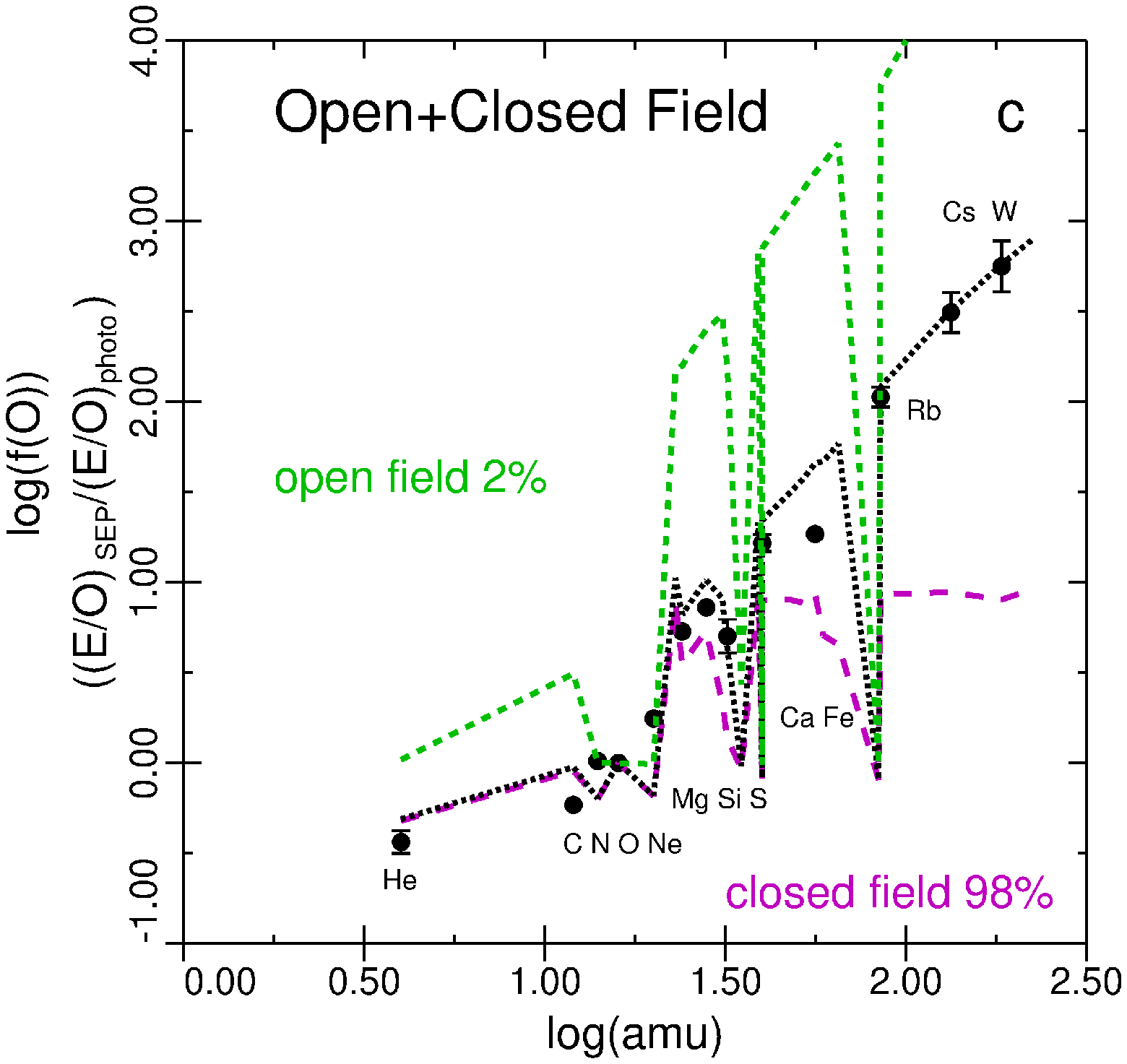}\includegraphics[width=3.5truein]{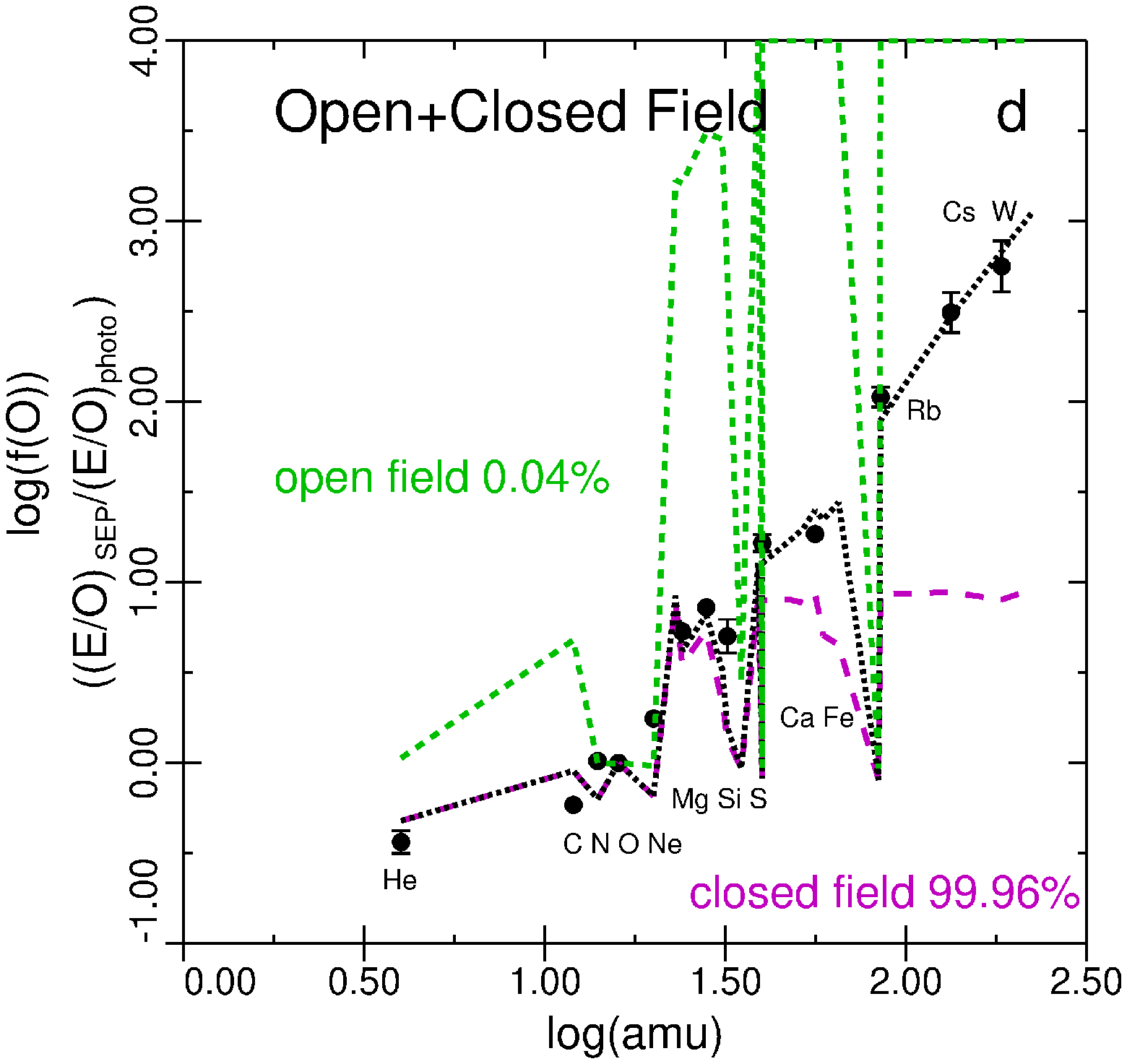}}
\vskip -0.75truein
\caption{Panel (a): Fractionations relative to O, f(O), defined as SEP element abundance divided by SEP O abundance, relative to the photospheric
elemental abundance relative to O, for torsional waves on closed fields with attack angles at the $\beta =6/5$ layer of 90$^{\circ}$ (magenta, long dash), 75$^{\circ}$ (green, short dash) and for shear waves at 90$^{\circ}$ (black, dotted). These are compared with data points (black circles with error bars) from \citet{mason07}. Panel (b): Similar models for open fields. The shear wave curve is the 
{\em maximum} fractionation possible for any wave amplitude.
Panel (c): Hybrid model (black dotted line), comprising 98\% fractionation 
from closed field model T0(h) (magenta long dash) and 2\% open field model T81(h). Panel (d): Hybrid model with 99.96\% from closed field model T0 (magenta long dash) and 0.04\% from open field model T90(h).}
\label{fig:IR}
\end{figure}

Table 1 and Figure \ref{fig:IR} illustrate preliminary calculations of such fractionations, compared
with data from \citet{mason07}, converted to fractionations based on photospheric abundances using data from \citet{caffau11,scott15a,scott15b} . 
The top left panel (a) of Figure \ref{fig:IR} shows two models for waves on a closed loop. Shear (black dotted curve) and torsional waves 
(magenta long dash curve) at an attack angle of 90$^{\circ}$ for
5 minute acoustic waves reaching the plasma $\beta =1.2$ layer are compared. In the closed loop case, there is little difference between them. We also show
torsional waves at an acoustic wave attack angle of 75 $^{\circ}$ (green short dash curve), which shows much lower fractionation than at 90$^{\circ}$. These are the closed field  models T90, S90, and T75 from Table 1, for torsional (T) and shear (S) waves. At 90$^{\circ}$ the light elements (He, C, N, O, Ne) are reasonably well matched. 
Intermediate mass (Mg, Si, S, Ca, Fe) are overpredicted while Rb, Cs, and W are about right. The models including
mode conversion show a general trend of increasing fractionation with increasing atomic mass, except for dips in the fractionation where a high FIP 
element is encountered. The model curves include Cl, Ar, and Kr, with atomic masses of 35, 40, and 84 amu respectively.

\begin{table}[ht]
\begin{center}
\caption{Closed and Open Field Fractionations}
\begin{tabular}{lccccccccccccr}
\hline
 && \multicolumn{4}{c}{closed Field}&&  \multicolumn{5}{c}{Open Field}&& Mason  \\

 & &T90& T75& S90& T0(h)&& T90& T75& S90& T81(h)& T90(h)&& (2007)\\
& $\delta v_{init}$ (km s$^{-1}$)&  0.48& 0.48& 0.48& 0.67& & 7& 7& 14& 20& 30& \\
ratio\\
\hline
He/O && 0.37& 0.60& 0.44& 0.48&& 1.08& 1.04& 1.07& 1.04&1.06&& 0.37 \\
C/O&  & 1.02& 1.03& 1.04& 0.90&& 1.38& 1.26& 1.70&  3.11&4.79&& 0.59\\
N/O& &  0.59& 0.73& 0.64& 0.63&& 0.99& 1.00& 0.99& 1.00&1.00&& 1.03\\
Ne/O&&  0.63& 0.75& 0.67& 0.66&& 0.97& 0.99& 0.97& 0.98&0.96&& 1.77\\
Mg/O&&  15.2& 2.49& 13.5& 3.53&& 4.56& 2.48& 9.78& 159.&1623.&& 5.35\\
Si/O& & 31.8& 3.12& 27.1& 5.33&& 5.48& 2.70& 12.1& 252.&3089.&& 7.27\\
S/O& &  2.90& 1.46& 2.74& 1.37&& 4.56& 2.51& 7.63& 152.&528.&& 5.03\\
Ca/O&&  85.3& 3.92& 70.8& 7.59&& 8.96& 3.34& 19.3& 712.&12036.&& 16.5.\\ 
Fe/O & &138.& 4.16& 133.& 8.64&& 16.4& 4.25& 30.7& 1858.&41065.&& 18.5.\\
Rb/O & &221.& 4.27& 180.& 8.63&& 43.8& 6.01& 55.3& 5624.&175470.&& 106.\\
Cs/O && 389.& 4.49& 314.& 8.80&& 171.& 9.08& 103.& 15596.&718723.&& 313.\\
W/O && 495.&  4.41& 401.& 8.03&& 554.& 12.3& 154.& 28506.&1683724.&& 563.\\

\hline \end{tabular}
\end{center}
\tablecomments{Fractionation relative to O given by equation 2. Starting wave velocity amplitude $\delta v_{init}$ given in the opposite chromosphere
for the closed field models, and at 500,000 km altitude for the open field models.}
\end{table}


The top right panel (b) shows the same set of models for the open field case. The shear wave 90$^{\circ}$ case shows the {\em maximum} fractionations
that can be found with shear waves. Lower and higher Alfv\'en wave amplitudes  and smaller attack angle all give lower fractionations, due to the interactions with sound waves.
It is clear that shear waves (black dotted curve) cannot account for the observations, and that torsional waves are necessary in the open field, as shown in the other two curves. 
The 90$^{\circ}$ torsional wave model (magenta long dash) matches the intermediate and high mass elements very well, but overpredicts He and C. As
expected, the 75$^{\circ}$ torsional waves (green short dash) do not match well at low or high mass. These models suggest that a combination of closed field
providing the low mass elements  (He, C, N, O, Ne) and open field supplying the high mass elements  (Rb, Cs, W) might give a good match for all element masses.
An important difference between the closed and open field cases
lies in the fractionation of C and S (and also P, included in the theoretical curves). These are technically high FIP elements, according to the definition that
their FIP is higher than the energy of a Lyman $\alpha$ photon, but these are the three high FIP elements with the lowest FIPs at 11.26, 10.487 and 10.360 eV
for C, P, and S respectively. These elements are known to be enhanced in abundance in accelerated particles sampled from co-rotating interaction regions
in the solar wind compared to gradual SEPs \citep{reames18}, and in the solar wind compared to the closed loop solar corona \citep{kuroda20,laming19}, and
that this difference is related to the fractionation region extending down to chromospheric altitudes where H is neutral \citep{laming19}.

The bottom left panel (c) shows a hybrid model. The closed field model with attack angle 0$^{\circ}$ (model T0(h) from Table 1)
is combined with a new open field torsional wave model with higher amplitude waves than the previous cases (T81(h) from Table 1), with an attack angle at $\beta = 6/5$ of 81.4$^{\circ}$. A proportion of 98\% closed field ions and 2\% open field ions gives
a very good match to the observational data given by \citet{mason07}, with only Fe being
mildly discrepant.  An attack angle of 0$^{\circ}$ in the closed gives a better match to the intermediate mass elements (Mg, Si, S, Ca, Fe) than an attack
angle of 90$^{\circ}$, but attack angles up to about 75$^{\circ}$ give similar results.
The closed field contribution dominates for the low mass elements, while the open
field dominates at higher element masses. Panel (d) shows a similar hybrid model with the same closed field fractionations, but now with the open field given
by model T90(h) in Table 1. The proportions are 99.96\% closed : 0.04\% open, and the model now gives a better account of the Fe fractionation, while
the other elements are still in reasonable agreement with the data.

\section{Discussion and Conclusions}

In all panels in Fig. 3 there is a trend of increasing fractionation with increasing element mass, with the exception of high FIP elements which remain unfractionated.
As one moves up the periodic table high FIP elements become increasingly rare, being restricted to the noble gases and Hg (FIP  = 10.438 eV). The high
FIP elements in our calculations are He, C, N, O, Ne, P, S, Cl, Ar, and Kr. It is clear from Table 1 that compared to the surrounding low FIP elements, S is
comparatively more fractionated in open field than in closed loops. The same is true for P, and this is visible in the various panels of Fig. 3. The minima
in fractionation close to atomic mass 40 amu corresponds to Cl (35 amu)  and Ar (40 amu); the minimum at 84 amu corresponds to Kr, with the low
FIP K appearing at  the maximum at 39 amu. The low mass side of the minimum can
be seen to be less steep for the closed field models compared to those for open field due to the relative depletion of P (31 amu) and S (32 amu). The ``kink'' in
the observational data around S strongly suggests that FIP plays a role in the fractionation. A similar feature is visible around S in the 2022 March 18-19
series of $^3$He-rich events observed by Solar Orbiter \citep{mason23}.

An important feature of the phenomenon is that the extreme fractionation seen in impulsive SEPs is not seen, or at least not so far detected, 
in the bulk solar wind. We argue that the fractionation is produced  in the chromosphere by the ponderomotive force of Alfv\'en waves generated by IR in the corona above. The Alfv\'en waves propagate down to the chromosphere and cause the fractionation before the electron heat
conduction arrives to initiate the chromospheric evaporation. This requires that the Alfv\'en speed be greater than the electron thermal speed, which
typically means a magnetic field greater than 100G. The evaporating material must move back up the same magnetic field lines that the Alfv\'en waves
came on, so that the fractionated plasma is injected straight back into the reconnection region and can be accelerated, without joining the quasi-thermal
solar wind.

We have assumed that these (non resonant) waves from reconnection affect mainly the chromosphere of the open field region. The chromosphere of
the closed field is fractionated by resonant waves, in the sense that the wave travel time from one loop footpoint to the the other is an integral number 
of wave half-periods. The low mass elements come from the closed field chromosphere, so that He and S are depleted relative to neighboring elements
as observed. The coronal loop behaves like a resonant cavity for Alfv\'en waves. The resonant waves are generated in the corona by nanoflares which 
represent small perturbations to the ambient magnetic field and ponderomotive force is restricted to the top of the chromosphere. By
contrast, the lack of resonance in the open field means that the ponderomotive force develops through the chromosphere, not just at the top, and the
strongest fractionation occurs lower down where the H is neutral. In these conditions, He is not depleted, and S fractionates more like other low
FIP elements. Further, interchange reconnection may not qualify as a ``small'' perturbation to the original magnetic field, and so any resonance that
may have existed would be lost.

Our model requires torsional Alfv\'en waves, because they couple much less strongly to sound waves that would otherwise destroy the mass dependent 
fractionation. The excitation of torsional Alfv\'en waves requires reconnection between open field and a twisted coronal loop, such that the twist is 
transferred to the open field after reconnection \citep[e.g.][]{kohutova20}. Shear waves generate too many sound waves. Circularly polarized waves would also be a possibility, but
these usually arise from wave-particle interactions at much higher frequency where the waves quickly damp through charge exchange and where the
ponderomotive force depends differently on atomic parameters. Otherwise, although a significant fraction of the reconnecting magnetic energy can
escape as MHD waves \citep[e.g.][]{kigure10,provornikova18} we have little guidance on what wave frequencies to expect. We might expect the distribution
of wave frequencies (or wavenumbers) produced to reflect the distribution of plasmoids, with in the case of nanoflare reconnection in a closed loop, the
loop resonant frequencies picked out by the boundary conditions. 

The apparent ubiquity of $^3$He events \citep{mason07,mason23} suggests trans-Fe elements may be enhanced in certain regimes of the solar wind. However 
the lack of correlation between $^3$He and heavy ion abundances in SEPs \citep{hart22} makes a quantitative estimate difficult. However the need for fractionation in both open and closed magnetic field strongly suggests a role for IR, and the need for torsional waves in the open field is consistent with the conditions surrounding coronal jets \citep[e.g.][]{mulay16,parenti21}.
Following the work of \citet{kazakov17,kazakov21}, we speculate that the $^3$He acceleration occurs in the open field region, where the $^4$He abundance is higher, and so this acceleration would be disconnected from the heavy ion SEP acceleration which is connected with the current sheet. There is much more work to be done along these lines. This is beyond the scope of this paper, where we have tried to establish the plausibility of fractionation based on the ponderomotive force as the mechanism behind the trans-Fe abundances in Impulsive SEP events. We emphasize that this is a model for the abundance enhancements only, and does not consider the acceleration, nor does it rule out any abundance modifications associated with the acceleration.
However it does offer a possibly more realistic means of enhancing trans-Fe abundances by the 
orders of magnitude required to match observations, compared with ``acceleration only'' scenarios.

\begin{acknowledgments}
This work has been supported by NASA HSR Grants NNH22OB102 and 80HQTR20T0076, and by Basic Research Funds of the Office of Naval Research. This study benefits from discussions within the International Space Science Institute (ISSI) Team ID 425 ``Origins of $^3$He-rich SEPs.''
\end{acknowledgments}



\end{document}